\documentclass[a4paper,12pt, english]{article}
\usepackage{graphicx}
\usepackage{hyperref}
\hypersetup{
    colorlinks=true,
    linkcolor=black,
    filecolor=black,      
    urlcolor=black,
    citecolor=black
}
\usepackage{cmap}
\usepackage{amssymb}
\usepackage{amsmath}
\usepackage{setspace}
\usepackage{indentfirst}
\usepackage{cite}

\usepackage{wrapfig}
\graphicspath{{figs/}}

\usepackage{bm}
\usepackage{siunitx}
\usepackage{booktabs}
\usepackage{array}
\usepackage{hyperref}
\hypersetup{
    colorlinks=true,
    linkcolor=blue,
    filecolor=magenta,      
    urlcolor=cyan,
    pdftitle={The Monte Carlo Method for the Orthonickelate Model},
    pdfpagemode=FullScreen,
}

\begin{document}



\title{\bf The Monte Carlo Method for the Orthonickelate Model}
\author{ Yu.D.~Panov$^1$, S.V.~Nuzhin$^1$, V.S.~Ryumshin$^1$, A.S.~Moskvin$^{1,2}$}
\date{}
\maketitle


{
$^1$Institute of Natural Sciences and Mathematics, Ural Federal University,
Yekaterinburg, Russia
}

{
$^2$M.N.~Mikheev Institute of Metal Physics, Ural Branch, Russian Academy of Sciences, Yekaterinburg, Russia
}

{
E-mail: 
}\href{mailto:nuzhin.stepan@urfu.ru}{nuzhin.stepan@urfu.ru}

\begin{center}
\textbf{ \Large Abstract}
\end{center} 

The peculiarities of phase states of the triplet boson model for orthonickelates are investigated analytically and by means of numerical simulations. 
The conditions of thermodynamic stability of homogeneous phases are found. 
It is shown that the description of the phase inhomogeneous state in the mean-field approximation qualitatively agrees with the observed state of the system in numerical simulations by the classical Monte Carlo method.


\newpage

\section{Introduction}
Due to their unique physical properties RNiO$_3$ orthoni ckelates (where R is rare earth or yttrium) are continuously studied with the use of increasingly advanced experimental methods~\cite{Medarde1997,Gawryluk2019}. This allows to clarify their phase diagram which includes (for various compounds) the metal-like non-ordered phase, charge ordered insulating phase, antiferromagnetic phase with noncollinear magnetic  structure resulting in such unique physical properties as metal-insulator transition. Some papers ~\cite{Gawryluk2019,Post2018,Mundet2021} reported discovery of a phase coexistence (or phase separation) in nickelates with Pr and Nd. In theoretical interpretation, there has been no common opinion so far regarding the formation mechanisms of orthonickelate’s electronic structure and phase diagrams.

We have reviewed earlier a model where  orthonickelates are considered as Jahn-Teller magnetic materials which are unstable with regard to anti-Jahn-Teller disproportionation reaction~\cite{Mazin2007} with formation of a system equivalent to the system of effective spin-triplet composite bosons moving in the non-magnetic  lattice~\cite{Moskvin2013,Moskvin2023ftt,Moskvin2023magchem,Ryumshim2024}. In the mean field approximation (MFA) for various values of the model nickelate’s parameters we have plotted the phase diagrams~\cite{Ryumshim2024} demonstrating the competition between the charge ordering phases, antiferromagnetic insulator and spin-triplet superconductor. Along with these phases, more complicated states were possible when several order parameters were non-zero. These states could be implemented as either homogeneous or in the form of phase separation. The well studied model of local singlet bosons~\cite{Micnas1990} with the use of numerical simulations demonstrated that a homogeneous phase akin a supersolid is metastable and unstable with respect to the phase separation~\cite{Batrouni2000}. Within MFA at finite temperatures for the triplet bosons model the phase separation also turns to be more stable compared to homogeneous phases ~\cite{Ryumshim2024}. This study was aimed at proving this suggestion  based on Maxwell phenomenological construction~\cite{Kapcia2012} using numerical simulation by classical Monte Carlo (MC) method allowing to kinematically account of constant  bosons concentration.

\section{Mean field approximation}\label{sec:MFA}

The general form of Hamiltonian for Jahn-Teller magnets where the orbital degeneracy is removed not due to Jahn-Teller effect, but due to  anti-Jahn-Teller disproportionation~\cite{Mazin2007,Moskvin2013} is given in papers~\cite{Moskvin2023magchem,Moskvin2023ftt}. For the rare-earth orthonickelates RNiO$_3$ the ion Ni$^{3+}$ in the low-spin  configuration t$_{2g}^{6}$e$_{g}^{1}$ of NiO$_6$ octahedron forms a Jahn-Teller center with ground state orbital doublet ${}^{2}$E. As a result of disproportionation, the electronic structure of the orthonickelate becomes a formal equivalent of the system of local composite spin-triplet bosons having configuration e$_{g}^{2}$;~${}^{3}$A$_{2g}$, which move in the non-magnetic lattice with  t$_{2g}^{6}$ centers. A simplified Hamiltonian for the model nickelate with a simple cubic lattice can be written as follows:
\begin{equation}
    \hat{H} = - t \sum_{\left\langle i j \right\rangle m} 
    \left( \hat{B}_{i}^{m +} \hat{B}_{j}^{m} + \hat{B}_{j}^{m +} \hat{B}_{i}^{m} \right)
    + V \sum_{\left\langle i j \right\rangle} \hat{n}_i \hat{n}_j 
    + J \sum_{\left\langle i j \right\rangle} \hat{\boldsymbol{\sigma}}_i \hat{\boldsymbol{\sigma}}_j. 
    \label{eq:H}
\end{equation}
Here, $t$ is a transfer integral of the spin-triplet boson with conservation of the spin projection $m=\pm1, 0$, $V$ is parameter of the inter-center charge-to-charge interaction, $\hat{n}_i$ is operator of the bosons number on $i$-th site, $J$ is exchange integral, $\hat{\boldsymbol{\sigma}}_i$ is operator of boson spin on i-th site. For bosons creation operators $\hat{B}_{i}^{m +}$ on $i$-th site in the state with projection of spin $m$ it is convenient to introduce Cartesian components with the help of relations $\hat{B}_{xi}^{m}=\frac{1}{2}\left(\hat{B}_{i}^{m+} + \hat{B}_{i}^{m}\right)$, 
$\hat{B}_{yi}^{m}=-\frac{i}{2}\left(\hat{B}_{i}^{m+} - \hat{B}_{i}^{m}\right)$, 
and use vector
$\hat{\mathbf{B}}_{i}^{m} = \big( \hat{B}_{xi}^{m} , \hat{B}_{yi}^{m} \big)$.
The explicit form of operators matrices is given in paper~\cite{Ryumshim2024}. It shall be noted that Hamiltonian~\eqref{eq:H} is a generalized version of a  well-know spinless hard-core bosons model~\cite{Micnas1990} for the spin-triplet
bosons case.

MFA for model~\eqref{eq:H} is reviewed in paper~\cite{Ryumshim2024}. The system
phase states are characterized by mean values
$\mathbf{B}_{\lambda(i)}^{m} = \big\langle \hat{\mathbf{B}}_{i}^{m} \big\rangle$, 
$n_{\lambda(i)} = \left\langle \hat{n}_{i} \right\rangle$ and  
$\boldsymbol{S}_{\lambda(i)} = \left\langle \hat{\boldsymbol{\sigma}}_{i} \right\rangle$,
where for the two mutually inter-penetrating lattices $A$ and $B$ of a simple cubic lattice we introduced the index $\lambda(i)$. Phase diagrams were plotted in variables $(T,n)$, where $T$ is temperature, $n=\big\langle \sum_{i} \hat{n}_i \big\rangle / N$ is the bosons concentration.

At high temperature the non-ordered (NO) phase, where  $\mathbf{B}_{\bar{\lambda}}^{m}=0$, $\mathbf{S}_{\lambda}=0$ and $n_{A}=n_{B}=n$, is activated.
If the temperature is quite low the solutions which can be called pure phases occur when only one order parameter is non-zero. This is the charge-ordered (CO) phase with  $x=\left(n_{A}-n_{B}\right)/2\neq0$,  antiferromagnetic (AFM) phase with $\mathbf{S}_{A}=-\mathbf{S}_{B}\neq0$ and boson superfluid (BS) phase with $\mathbf{B}_{\lambda}^{m}\neq0$. Also, under certain conditions, solutions  for mixed phases can be implemented, when several order parameters are non-zero, e.g., phase akin supersolid in the local bosons model~\cite{Micnas1990}. However, as with local bosons model~\cite{Kapcia2013}, free energy of mixed phases in model~\eqref{eq:H} in MFA is higher than the free energy of the pure phases separation state. The phase separation is set by Maxwell construction~\cite{Kapcia2012}: 
at a given temperature the boundary concentrations $n_{i}$ corresponding to pure phases  $i=1,2$ can be found from the concentration ratios of chemical potential of phases:$\mu_{i}(n,T)=\mu^{\ast}$, where $\mu^{\ast}$ is the point of intersection of specific grand potentials of  phases,$\omega_{1}(\mu^{\ast},T)=\omega_{2}(\mu^{\ast},T)$. 
Free energy has the following form
\begin{equation}
    f(n, T) = m_1 \, f_1(n_1,T) + m_2 \, f_2(n_2,T),
    \label{eq:fPS}
\end{equation}
where  $f_{i}$ are pure phases specific free energies calculated at boundary concentrations $n_{i}$, $n_{1} \le n \le n_{2}$, and the volume fractions $m_{i}$ of phases are set by the ratios $m_{1}=(n_{2}-n)/(n_{2}-n_{1})$, $m_{2}=(n-n_{1})/(n_{2}-n_{1})$.

It should be noted that the Maxwell construction~\eqref{eq:fPS} is
a hypothesis that looks reasonable from the physical standpoint, however, it needs an independent confirmation. 
For the local spinless bosons the authors have demonstrated~\cite{Batrouni2000} the thermodynamic instability of the homogeneous phase and correctness of the phase separation suggestion using numerical simulation by the quantum MC method. 
In this study we use classical MC method which allows a direct observation of the current system state.
For clarity, we’ll consider a 2D square lattice ($z=4$).

\begin{figure}
	\centering
	\includegraphics[width=0.5\textwidth]{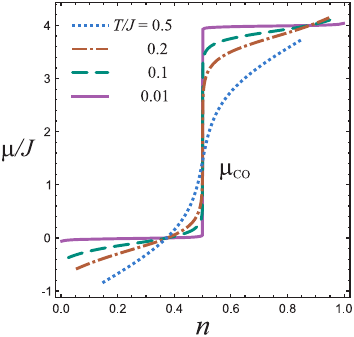} 
	\caption{Chemical potential of CO phase $z=4$, $V/J=1$ for various $T$. Boundary concentrations are defined by the ratio $T=4Vn(1-n)$.}
	\label{fig:muCO}
\end{figure}

Since one of the parameters of the simulated system
is the bosons concentration, it is important to consider
the concentration dependence of chemical potential  $\mu$ for
different phases. Figure~\ref{fig:muCO} illustrates the dependences of CO
phase chemical potential in MFA for several temperature
values.
The boundary concentrations are defined by
expression for critical temperature of the charge ordering,
$T_{CO} = zVn \left( 1-n \right)$, which coincides with the local bosons model case ~\cite{Micnas1990}. Condition of phase stability corresponds
to positive value of derivative $\left( \frac{\partial\mu}{\partial n} \right)_T$.
As seen from Figure~\ref{fig:muCO}, the CO phase is stable at all finite tempera-
tures, however, for low temperatures the stability condition worsens, $\left( \frac{\partial\mu}{\partial n} \right)_T\to0$, for all concentrations except for
$n=0.5$.
In terms of selecting the MC algorithm option for numerical simulation, the flat sections of $\left( \frac{\partial\mu}{\partial n} \right)_T$ dependence make it difficult to carry out the calculations
within a grand canonical ensemble:
at sufficiently low temperatures it is required to significantly increase both, the $\mu$ partition, and the number of MC steps.

\begin{figure}
	\centering
	\includegraphics[width=0.45\textwidth]{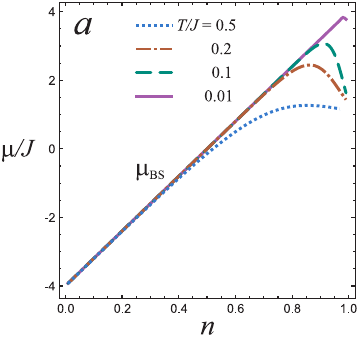} 
	\quad
	\includegraphics[width=0.45\textwidth]{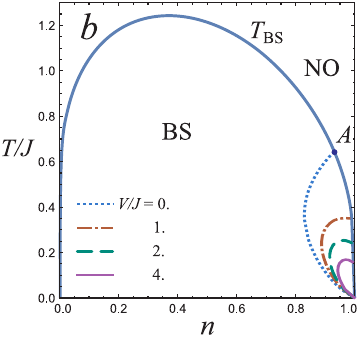} 
	\caption{
        \textit{a} is the chemical potential of BS phase at $z=4$, $t/J=1$, $V=0$ for different $T$. 
        \textit{b} are the boundaries of BS phase stability at different values $V$ ($z=4$, $t/J=1$). 
        Critical temperature $T_{BS} = 4t \left(  4n-3 \right) \left[ 3 \ln\frac{n}{3\left( 1-n \right) } \right]^{-1}$.
        }
	\label{fig:muBS}
\end{figure}

Concentration dependencies of chemical potential of BS phase in MFA are shown in Figure.~\ref{fig:muBS} at $V=0$. In $n<0.5$ area they can be expressed by formula $\mu=z\left[ Vn + t\left( 2n-1 \right) \right]$ for the local bosons model. When $n>0.5$ the ratios of BS phase chemical potential versus $n$ are non-linear and can have sections with $\left( \frac{\partial\mu}{\partial n} \right)_T<0$. The boundaries of respective BS phase stability areas are shown in Figure~\ref{fig:muBS}. The states in the area between the stability boundary at a given $V$ and curve for BS phase critical temperature $T_{BS} = zt \left(  4n-3 \right) \left[ 3 \ln\frac{n}{3\left( 1-n \right) } \right]^{-1} $ on the right will correspond to the phase separation into BS and NO phase macroscopic domains.
 The tricritical point $A$ divides the curve $T_{BS}$ into transition lines of 2-d order on the left and 1-st order on the right of $A$.
 Position of $A$ point is shifted to $n=1$, and the size of phase separation area on phase diagram is decreased with the growth of $V$.

\begin{figure}
	\centering
	\includegraphics[width=0.45\textwidth]{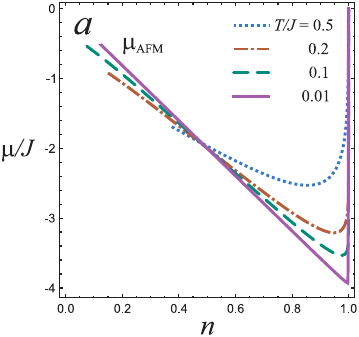} 
	\quad
	\includegraphics[width=0.45\textwidth]{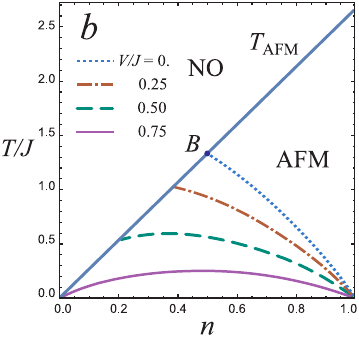} 
	\caption{
        \textit{a} — chemical potential of AFM phase at $z=4$, $J=1$, $V=0$ for various $T$. 
        \textit{b} are the boundaries of AFM phase stability at different values$V$ ($z=4$, $J=1$). 
        Critical temperature $T_{AFM} = 8Jn/3$.
        }
	\label{fig:muAFM}
\end{figure}

Instability of homogeneous AFM phase in MFA at low
values $V/J$ is shown in Figure~\ref{fig:muAFM}. The dependencies on the
left panel of Figure~\ref{fig:muAFM} for the chemical potential versus $n$ at
$V=0$ have sections with negative derivative at rather low $T$.
The boundaries of stability of homogeneous AFM phase are shown in the right panel in Figure~\ref{fig:muAFM} for several values $V/J$.
Homogeneous AFM phase is stable in the area to
the right and above these boundaries and below critical
temperature $T_{AFM} = 2zJn/3$, and in the area to the left and
below these boundaries and below $T_{AFM}$ a phase separation
occurs with separation into macroscopic domains of AFM
and NO phases. The tricritical point $B$ divides the curve
$T_{AFM}$ into transition lines of 1-st order on the left of and
2-d order on the right of $B$. Position of B reaches $n=0$
at  $V/J=0.75$, and at $V/J>1$ homogeneous AFM phase
becomes stable at $T<T_{AFM}$ at all  $n$.

\section{Boson concentration condition taken into account in classical Monte Carlo  method}\label{sec:MC}

The nature of chemical potential dependencies of various phase states of the model indicates the complexity of its numerical simulation by MC method within formalism of a grand canonical ensemble.
Further, we’ll consider a classic algorithm with a kinematic accounting of the bosons concentration.
One of the benefits of this algorithm is a possibility to visualize the evolution of the lattice states (averaged in MC steps) with a temperature at specified
concentration.

Let’s describe the system state using quasi-classical wave
function $\left| \Psi \right\rangle = \prod_{i}\left| \psi_{i} \right\rangle$, where for the on-site wave function
\begin{equation}
	\left| \psi_{i} \right\rangle 
	= c_{1,11}^{(i)} \left| 1 , 1 1 \right\rangle_{i} 
	+ c_{1,10}^{(i)} \left| 1 , 1 0 \right\rangle_{i} 
	+ c_{1,1{-}1}^{(i)} \left| 1 , 1 \, {-}1 \right\rangle_{i}
	+ c_{0,00}^{(i)} \left| 0 , 0 0 \right\rangle_{i}
	\label{eq:psi}
\end{equation}
In the wave functions $\left| n,Sm \right\rangle$ in formula~\ref{eq:psi} $n$ is the number of bosons, $S$ is the spin value, $m$ is the value of $z$-projection of the spin.
Coefficients $c_{n,Sm}^{(i)} = r_{n,Sm}^{(i)} \, e^{i\phi_{n,Sm}^{(i)}}$, $\phi_{n,Sm}^{(i)}\in[0,2\pi]$ are normalized to 1. This allows selecting their parametrization as follows:

\begin{eqnarray}
	r_{1,11}^{(i)} &=&  \sin\theta_{i} \sin\psi_{i} \cos\varphi_{i} , \label{eq:r11} \\
	r_{1,10}^{(i)} &=&  \sin\theta_{i} \cos\psi_{i} , \label{eq:r10} \\
	r_{1,1{-}1}^{(i)} &=&  \sin\theta_{i} \sin\psi_{i} \sin\varphi_{i} , \label{eq:r1m1} \\
	r_{0,00}^{(i)} &=&  \cos\theta_{i} , \label{eq:r00} 
\end{eqnarray}
where $\theta_{i},\psi_{i},\varphi_{i} \in [0,\frac{\pi}{2}]$.
Homogeneous sampling in the
state space on the site corresponds to a uniform distribution of points on the surface of a unit sphere in 8D coordinate space $\left( x_{n,Sm}^{(i)}, y_{n,Sm}^{(i)} \right)$, where $c_{n,Sm}^{(i)} = x_{n,Sm}^{(i)} + i y_{n,Sm}^{(i)}$. This can be achieved by generation of 7 random values: 
$\alpha=\sin^6\theta_{i}$, $\beta=\sin^4\psi_{i}$, $\gamma=\sin^2\varphi_{i}$ and $\xi_{n,Sm}=\phi_{n,Sm}^{(i)}/2\pi$,
that are uniformly distributed in section  $[0,1]$.

Since bosons have the on-site boson density $n_{i}=\left\langle \hat{n}_{i} \right\rangle= \sin^2\theta_{i} = \alpha^{1/3}$, the distribution function $F(n_{i})$ and corresponding probability density are expressed as
\begin{equation}
	F(n_{i}) = \int\limits_0^{n_{i}^3} \, d\alpha = n_{i}^3 , \quad 
	f(n_{i}) = 3 n_{i}^2. 
	\label{eq:f(n)}
\end{equation}
For a pair of sites the probability density that the first site has bosons density of $n_1$, the second site has density of $n_2$, is equal $f(n_1)f(n_2)$, since the states in different sites we consider as independent variables. From the sections of
function $f(n_1)f(n_2)$, that correspond to a mean density on the pair of sites $\bar{n}=\left( n_1+n_2 \right) /2$, the probability densities and distribution functions for $n_1$ with the set value $\bar{n}$ can be obtained:
\begin{equation}
	f_1(n_1;\bar{n}) = \frac{n_1^2 (2\bar{n}-n_1)^2}{\displaystyle \int\limits_{n_{1,min}}^{n_{1,max}} \!\!\! x^2 (2\bar{n}-x)^2 \, dx}
	\label{eq:f1(n)}
\end{equation}
\begin{equation}
	F_1(n_1;\bar{n}) = \displaystyle \int\limits_{n_{1,min}}^{n_{1}} \!\!\! f_1(x;\bar{n}) \, dx
	= \;\; \frac{\varphi(n_1) - \varphi(n_{1,min})}{\varphi(n_{1,max}) - \varphi(n_{1,min})},
	\label{eq:F1}
\end{equation}
where 
\begin{equation}
	\varphi(x) = \frac{4 n^2 x^3}{3}-n x^4+\frac{x^5}{5} , 
\end{equation}
$n_{1,min}$ and $n_{1,max}$ are the limits of a range where density can
variate on one of the sites at a set value $\bar{n}$: 
\begin{equation}
	n_{1,min} = \bar{n}-\tfrac{1}{2} + |\tfrac{1}{2}-\bar{n}| , \quad
	n_{1,max} = \bar{n}+\tfrac{1}{2} - |\tfrac{1}{2}-\bar{n}| .
\end{equation}

As a result, the algorithm for selecting the states on a pair of sites that provides on each step of MC the constant bosons density in the system can be formulated as follows.
\begin{enumerate}
	
	\item For the selected pair of sites we need to define the mean density: $\bar{n} = \left(n_{1,0} + n_{2,0}\right)/2$. 
	
	\item Let’s find the new value $n_1$ on one of the sites
    from the equation $F_1(n_1;\bar{n}) = \gamma$,
    where $F_1$ is defined by expression~\eqref{eq:F1}, $\gamma$ is a random variable uniformly distributed in $[0,1]$.
	
	\item Let’s find the new value $n_2$ on the second site: $n_2=2\bar{n}-n_1$.

	\item Let’s find the values $\theta_i = \arcsin\left(\sqrt{n_i}\right)$, $i=1,2$.
	
	\item Now we need to find $\psi_i = \arcsin\left(\sqrt[4]{\beta_i}\right)$ and $\phi_i = \arcsin\left(\sqrt{\gamma_i}\right)$, $i=1,2$, where $\beta_i$ and $\gamma_i$ are random variables uniformly distributed in $[0,1]$.
	
	\item Let’s calculate $r_{n,Sm}^{(i)}$ by formulas~(\ref{eq:r11}--\ref{eq:r00}).
	
	\item Now we need to generate the uniformly distributed ran-
    dom phases $0 \leq \phi_{n,Sm}^{(i)} \leq 2\pi$, $i=1,2$, and find new coefficients of the wave functions on the sites: 
    $c_{n,Sm}^{(i)} = r_{n,Sm}^{(i)} \, e^{i\phi_{n,Sm}^{(i)}}$.
	
\end{enumerate}
When changing the system states we use standard Metropolis algorithm of MC method.

\section{Results}\label{sec:results}

\begin{figure}
	\centering
	\includegraphics[width=0.6\textwidth]{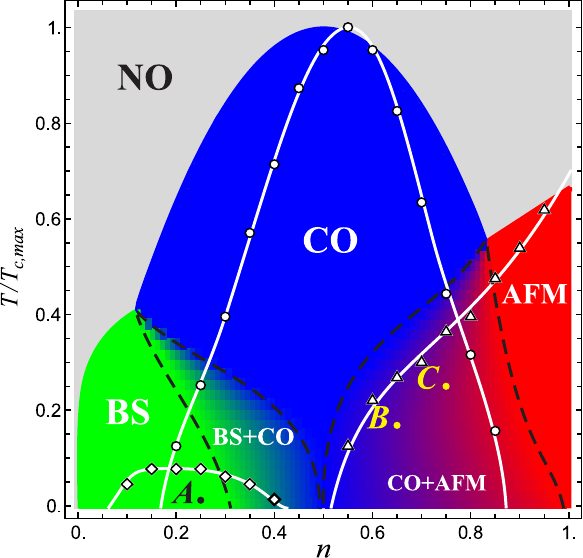} 
	\caption{
        Comparison of phase diagrams, obtained in MFA and by MC method for the lattice of $N=96\times96$ sites at $z=4$, $V/J=4$, $t/J=1.5$.
        The temperature scale is set for MFA $T_{c,max}^{(1)}=4J$, for MC method $T_{c,max}^{(2)}=0.63J$. 
        Homogeneous colors indicate the areas of different phases of MFA, dash lines are the boundaries of the phase separation areas in MFA.
        Rhombi, circles and triangles show the points, where values calculated by MC
        method for the order parameters for BS, CO and AFM phase reach the threshold value of $1\%$ of the maximal possible value.
        Points $A$, $B$ and $C$ correspond to the system states shown in
        Figure~\ref{fig:lattice-states}.
	}
	\label{fig:pd-MFA-MC}
\end{figure}

Figure~\ref{fig:pd-MFA-MC} gives comparison of the phase diagram in MFA and results of simulation by classical MC method allowing to provide the boson concentration constancy kinematically.
The temperature scale is set by maximal critical temperature that has sufficiently different values for MFA and MC method.
With selected parameters  $z=4$, $V/J=4$, $t/J=1.5$ for MFA the maximal temperature
will be the critical temperature of CO phase at $n=0.5$: $T_{c,max}^{(1)} = 4J$. Areas of homogeneous colors indicate different phases of MFA, dash lines designate the binodals. At $0.15<n<0.50$ there’s an area of phase separation for BS and CO phases, and at $0.5<n<1.0$ there’s for CO and AFM phases.
 
Numerical simulation was carried out for a square lattice of $N=96\times96$ sites with periodic boundary conditions within temperature $T/J$ range from 0.01 to 0.80 with a step 0.01 for bosons concentrations, from 0.05 to 0.95 with a step 0.05. 
For every temperature value the $4\cdot10^{6}$ steps of MC were fulfilled. 
As order parameters we calculated the mean values as follows: for CO phase $x=\left| \sum_{i} (-1)^{i} n_{i} \right| / 2N$, where $n_{i}$ is the on-site boson concentration, $(-1)^{i}=\pm1$ is for various sublattices, for BS phase $B=\left| \sum_{i} \mathbf{B}_{i} \right| / N$, for AFM phase $L=\left| \sum_{i} (-1)^{i} \mathbf{S}_{i} \right| / N$.

\begin{figure}
	\centering
	\includegraphics[width=0.45\textwidth]{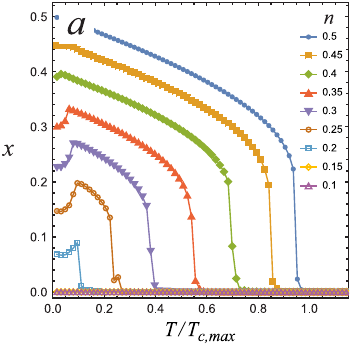} 
	\qquad
	\includegraphics[width=0.45\textwidth]{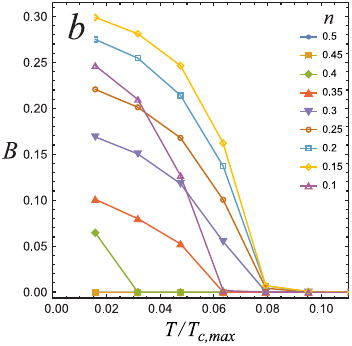} 
	\caption{
        Temperature dependencies of order parameters for
        CO phase (a), BS phase (b) for concentrations $n=0.1 \text{---} 0.5$.
		}
	\label{fig:op-CO-BS}
\end{figure}

\begin{figure}
	\centering
	\includegraphics[width=0.45\textwidth]{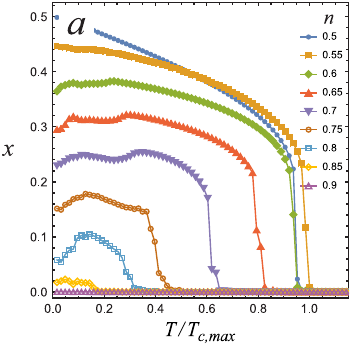} 
	\qquad
	\includegraphics[width=0.45\textwidth]{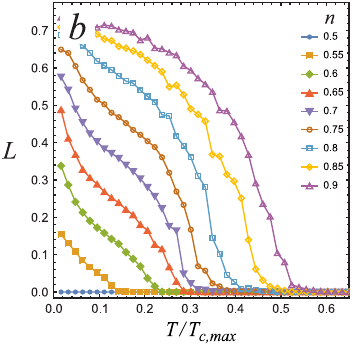} 
	\caption{
        Temperature dependencies of order parameters for
        CO phase (a), AFM phase (b) for concentrations $n=0.5\text{---}0.9$.
		}
	\label{fig:op-CO-AFM}
\end{figure}

Temperature dependencies of order parameters for different concentrations are given in Figures~\ref{fig:op-CO-BS} and~\ref{fig:op-CO-AFM}.
To estimate the critical temperature we used a value with which the corresponding order parameter reached the threshold value of 1\% of the maximal possible value. With selected parameters $T_{c,max}^{(2)}=0.63J$ is obtained at $n=0.55$ for CO phase.
This value was used as a temperature scale in Figure~\ref{fig:op-CO-BS},\ref{fig:op-CO-AFM} and~\ref{fig:op-CO-AFM} and in Figure~\ref{fig:pd-MFA-MC} for this MC calculations, where the corresponding points for CO phase are designated by circles, for BS phase by rhombi, for AFM phase by triangles. 
The white line connecting these points is given for graphic representation only.
Though the estimate of critical temperature is sensitive to the threshold value, the view of the phase diagram $T/T_{c,max}$ generally remains the same. 
More extensive analyses with the estimate of critical temperatures as per scaling theory are planned to be made in the future.

Comparison of the phase diagrams obtained through MFA and MC method demonstrates that, though the ratio of the temperature scales is $T_{c,max}^{(1)}/T_{c,max}^{(2)}\simeq6.3$, the ratio between maximal critical temperatures of CO and AFM phases persists. 
At that, for BS phase this ratio turns out to be significantly less than in MFA.
Also we can see relative reduction of areas of the ordered phases in the phase
diagram, obtained by MC method compared to MFA, and the shift of the critical temperature maximum of CO phase towards the point $n=0.55$.

\begin{figure}
	\centering
	\includegraphics[width=0.3\textwidth]{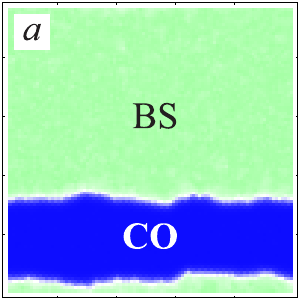} 
	\quad
	\includegraphics[width=0.3\textwidth]{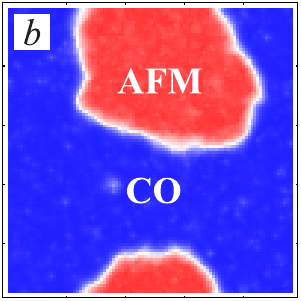} 
	\quad
	\includegraphics[width=0.3\textwidth]{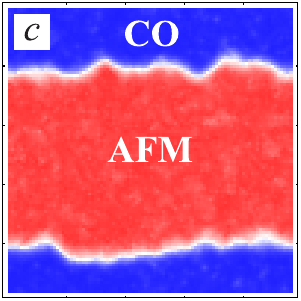} 
	\caption{
        Lattice state from $N=96\times96$ sites for $z=4$, $V/J=4$, $t/J=1.5$ at 
        a --- $n=0.25$, $T/T_{c,max}=0.02$ (point $A$ in Figure.~\ref{fig:pd-MFA-MC});
        b --- $n=0.65$, $T/T_{c,max}=0.17$ (point $B$ in  Figure.~\ref{fig:pd-MFA-MC});
        c --- $n=0.75$, $T/T_{c,max}=0.25$ (point $C$ in Figure.~\ref{fig:pd-MFA-MC}).
		}
	\label{fig:lattice-states}
\end{figure}

Temperature dependencies of CO phase order parameter in Figure~\ref{fig:op-CO-BS}, a at $n=0.2\text{---}0.4$ have an inflection in those points where a non-zero value of BS phase order parameter appears in Figure~\ref{fig:op-CO-BS}, b. 
The system state at $n=0.25$ and $T/T_{c,max}=0.02$ is shown in Figure~\ref{fig:lattice-states}, a.
The areas with the non-zero mean local value of order parameter for only one of the phases are highlighted by different colors. 
Thus, phase separation of BS and CO phases correspond to the system state in point $A$ in Figure~\ref{fig:pd-MFA-MC}.

The quantities $x$ and $L$ in Figure~\ref{fig:op-CO-AFM} for $n=0.55\text{---}0.80$ also have temperature intervals where both order parameters are non-zero. 
The phase states at $n=0.65$ and  $T/T_{c,max}=0.17$ and at $n=0.75$ and $T/T_{c,max}=0.25$, which corresponds to the points $B$ and $C$ in Figure~\ref{fig:pd-MFA-MC}, are shown in Figure~\ref{fig:lattice-states}, b and c. In both cases the phase separation into macroscopic areas of CO and AFM phase takes place, and in $C$ point the fraction of AFM is greater than in $B$, which is well consistent with MFA results.

\section{Conclusion}\label{sec:conclusion}

The peculiarities of the triplet bosons model phase states for the orthonickelates were investigated by MFA and using numerical simulation via MC method.
It was demonstrated that, in contrast to the singlet local bosons model~\cite{Micnas1990}, the BS phase of triplet bosons was unstable at high concentrations relative to the  phase separation with a non-ordered phase. Also, AFM phase is unstable at small inter-center charge-to-charge interaction, $V/J<1$. 
The numerical simulation by classical MC method demonstrated that instead of phases with several nonzero order parameters a phase separation takes place.
This is well consistent with MFA results~\cite{Ryumshim2024}, where Maxwell construction was used to analyze the thermodynamic properties of the phase-heterogeneous state.

\subsection*{Funding}
This study was supported by the project FEUZ-2023-0017 of the Ministry of Education and Science of the Russian Federation.



\end{document}